\documentclass[preprint,showpacs,aps,nofootinbib,superscriptaddress]{revtex4-1}
\usepackage{amssymb}
\usepackage{amsmath,amssymb,graphicx}
\usepackage{graphicx}
\usepackage{dcolumn}
\usepackage{bm}
\usepackage{multirow}

\def\beq{\begin{equation}}
\def\eeq{\end{equation}}
\def\bea{\begin{eqnarray}}
\def\eea{\end{eqnarray}}

\usepackage{pstricks}
\usepackage{graphicx}
\usepackage{amsbsy}
\usepackage{bm}
\usepackage{fancyhdr}
\usepackage{epsfig}
\usepackage{slashed}

\begin{document}

\bigskip

\vspace{2cm}
\title{
Long-distance weak annihilation contribution to the $B^{\pm}\to (\pi^{\pm},K^{\pm}) \ell^+\ell^-$ decays}
\vskip 6ex
\author{A. Guevara}
\email{aguevara@fis.cinvestav.mx}
\affiliation{Departamento de F\'isica, Centro de Investigaci\'on y de Estudios Avanzados del Instituto Polit\'ecnico Nacional, 
Apartado Postal 14-740, 07000 M\'exico D.F., M\'exico}
\author{G. L\'opez Castro}
\email{glopez@fis.cinvestav.mx}
\affiliation{Departamento de F\'isica, Centro de Investigaci\'on y de Estudios Avanzados del Instituto Polit\'ecnico Nacional, 
Apartado Postal 14-740, 07000 M\'exico D.F., M\'exico}
\author{P. Roig}
\email{proig@fis.cinvestav.mx}
\affiliation{Departamento de F\'isica, Centro de Investigaci\'on y de Estudios Avanzados del Instituto Polit\'ecnico Nacional, 
Apartado Postal 14-740, 07000 M\'exico D.F., M\'exico}
\author{S. L. Tostado}
\email{stostado@fis.cinvestav.mx}
\affiliation{Departamento de F\'isica, Centro de Investigaci\'on y de Estudios Avanzados del Instituto Polit\'ecnico Nacional, 
Apartado Postal 14-740, 07000 M\'exico D.F., M\'exico}

\bigskip
\begin{abstract}
We propose an alternative evaluation of the long-distance weak annihilation (WA, also called one-photon exchange in this paper) contribution to the rare semileptonic $B^{\pm}\to (\pi^{\pm},K^{\pm})\ell^+\ell^-$ ($\ell=e,\,\mu$) decays. This hadronic description at low energies is matched at intermediate energies to its short-distance counterpart in terms of quark and gluon degrees of freedom.  
 Although the WA contribution does not contribute to solve the possible breaking of lepton-universality observed by LHCb in the $B^{\pm}\to K^{\pm}(\mu^+\mu^-/e^+e^-)$ ratio, nor provides an important hadronic contamination to their decay rates, its contribution to the branching ratios (and direct CP asymmetry) of the $B^{\pm}\to \pi^{\pm}\ell^+\ell^-$ transitions turns out to be significant. This hadronic pollution should be taken into account when looking for new physics effects in decays into pions, which suggests to restrict these searches to squared lepton-pair invariant mass in the $(1,8)$ GeV$^2$ range. The interference of the one-photon exchange contribution with the dominant short-distance one-loop amplitude induces a sizable CP asymmetry in these rare decays, which calls for dedicated measurements.
\end{abstract}

\pacs{13.20.He, 12.38.Lg, 13.40.Gp, 11.30.Er}
\keywords{B-Physics, rare decays, Electromagnetic form factors, CP violation}
\maketitle
\bigskip

\section{Introduction}
Rare meson decays are expected to serve as harbinger for New Physics (NP) in experiments at the intensity frontier 
\cite{Buchalla:1995vs,rarekreview,rarebreview}. Better measurements or upper limits on a plethora of rare (semi)leptonic and radiative $K$ and $B$ meson decays 
in forthcoming experiments, compared to precise Standard Model (SM) predictions, eventually  will provide indirect indications of heavier particles with new 
interactions. Particularly sensitive for NP searches are those decays dominated by short-distance (SD) dynamics where the hadronic uncertainties are well under 
control. Conversely, precise measurements of these rare decays, combined with non-observation of NP effects, will furnish a better determination of flavor mixing parameters.

Very recently, the LHCb collaboration has reported a deficit in the ratio of muon to electron pairs produced in 
$B^\pm \to K^\pm \ell^+\ell^-$ decays, in the $\left(1,6\right)$ GeV$^2$ region for the squared invariant-mass of the lepton pair 
\cite{lhcb}. This energy region is cleverly chosen as it excludes long-distance (LD) contributions associated to charmonium and light
 vector resonance production in $B^-$ decays followed by their conversion to lepton pairs (for definiteness, hereafter we will focus on decays of negatively charged $B$ mesons, unless otherwise indicated). The measured ratio 
$R_K\equiv B(B^-\to K^-\mu^+\mu^-)/B(B^-\to K^-e^+e^-)=(0.745^{+0.090}_{-0.074}\pm0.036)$ \cite{lhcb}, if confirmed in more refined
 measurements, would call for lepton universality violating interactions since in the SM $R_K=1.0003\pm0.0001$ 
\cite{bobeth, bobeth2, hiller1, hiller2} for the energy region reported by LHCb \cite{lhcb}. Possible explanations involving NP interactions 
 have been suggested as the source of non-universal leptonic interactions \cite{newphysics}. Interestingly, other
 anomalies have been reported in angular observables of related $B\to K^*\ell^+\ell^-$ decays \cite{btokstar-exp} which have been widely discussed in the 
 literature \cite{btokstar1}.

In the SM, the semileptonic $B^- \to P^-\ell^+\ell^-$ ($P=K,\pi$) decays are dominated by the SD $b\to (s,d) \ell^+\ell^-$  transition 
\cite{bobeth, bobeth2, hiller1, hiller2, Khod, btopi0, btopi, ali}. This elementary process is induced by the electromagnetic and weak penguin 
 (Figure 1a), as well as $W$ boson box (Figure 1b) diagram contributions, which are dominated by loops involving the top quark.  In these exclusive processes, LD 
 chiral corrections to the hadronic matrix element have been computed for soft momenta of light pseudoscalars (i.e. high-momentum $q$ of dilepton pairs), where they amount to 
 rather large $\mathcal{O}(20,30\%)$ corrections \cite{otherend}. Our point here is that, although this kind of corrections will be much smaller on the other 
 energy end (large recoil region for the heavy-to-light B-meson form factors \cite{largereco}), the high precision of measurements as well as the sharp 
 predictions of the dominating SD contributions suggests LD effects associated to resonance dynamics at the GeV scale be taken into account as well for these more energetic $P$ mesons. Contributions of four-quark operators giving rise to tree-level WA amplitudes with virtual photon emission are also possible for exclusive $B^- \to P^-\ell^+\ell^-$ processes and are the main concern of this paper.

 As far as we know, the one-photon exchange Feynman diagrams (arising from WA operators) shown in Figures 1c-1d provide a so far neglected LD contribution that must be included 
 in the calculation of these exclusive semileptonic rare decays \footnote{We neglect a subdominant
$O(G_F^2)$ neutrino-exchange amplitude. This one was checked to give a highly suppressed effect in $K^{\pm} \to \pi^{\pm}\ell^+\ell^-$ decays \cite{nu-exchange}.}. These 
 additional LD contributions could in principle modify the $R_P$ value due to kinematical effects \cite{Guevara:2013wwa}. According to chiral effective theories of QCD 
\cite{ChPT}, for photon four-momenta below $\sim2$ GeV, 
photon-hadron interactions are best formulated in terms of the relevant (pseudo-)scalar and (axial-)vector hadronic degrees of freedom. As it was shown long ago \cite{ecker}, the  $K^{\pm} \to \pi^{\pm}\ell^+\ell^-$ rare semileptonic decays 
 are dominated by the one-photon exchange contributions analogous to the ones of Figure 1c and 1d; the SD top-quark loop penguin and W-box contributions in that case are negligible small \cite{Buchalla:1995vs}. Owing to gauge-invariance, this LD contribution vanishes at lowest order in chiral perturbation theory but this is not the case at higher orders \cite{ecker, otherkl3}. 

According to the respective CKM mixing factors in the SD and LD amplitudes of $B^- \to P^-\ell^+\ell^-$ decays, one can envisage a correction at the 
 percent level for $P=K$ but a slightly higher effect could be expected for $P=\pi$ depending on the limits of the integrated observables. It is worth noticing that WA contributions were previously considered for $B\to (K^*/\rho)\ell^+\ell^-$ \cite{beneke} and $B\to \pi \ell^+\ell^-$ \cite{Hou:2014dza} decays in the framework of the QCD factorization (QCDf) scheme \cite{QCDf}. 
Recently, non-local effects in $B \to \pi \ell^+\ell^-$ decays have been computed in Ref.~\cite{Hambrock:2015wka} using the method developed in Ref.~\cite{Khodjamirian:2010vf} and previously used in Ref.~\cite{Khod} 
for $B \to K \ell^{+}\ell^{-}$ decays for low dilepton invariant masses. 
At large hadronic recoil (corresponding to low values of the squared virtual photon momentum $q^2\ll M_B^2$), the light-meson energy ($E_P$) is much larger than the energy scale of hadronic binding: $E_P\gg \Lambda_{QCD}$. Then, the virtual photon exchange between the hadrons and the dilepton pair and hard gluon scattering can be systematically expanded in powers of $1/E_P$ using QCDf. However, the $B \to P \gamma$ amplitude suffers from nonperturbative corrections due to $u\bar{u}$ and $c\bar{c}$ intermediate states, which form the $\rho$, $\omega$, $J/\Psi$, ... resonances \cite{Lim:1988yu} (similarly, in $D\to P\gamma$ processes $d\bar{d}$ and $s\bar{s}$ can be excited and the $\phi$ meson plays a prominent role \cite{Burdman:2001tf}). Therefore, a clean QCDf prediction shall be 
limited to an energy range which excludes these resonant contributions, conservatively $2\leq q^2\leq 6$ GeV$^2$.

On the contrary, the extension of these limits requires the inclusion of near-threshold $u\bar{u}$ and $c\bar{c}$ ($d\bar{d}$ and $s\bar{s}$) resonances. Let us recall Ref.~\cite{Lim:1988yu} where these are accounted for studying the $b \to s \gamma$ processes leading to a relevant effect of the charmonium resonances in the inclusive decay width. Also, Ref.~\cite{Burdman:2001tf}  
considered resonance effects from $q\bar{q}$ excitations in $D\to X_u \ell^+ \ell^-$ processes, where $\phi$ exchange basically saturates the decay width. In the analyzed $B^-\to P^-\ell^+\ell^-$ decays, cutting at a maximum $q^2$ value well below 
the charmonium region kills the effects of the $c\bar{c}$ resonances and we have checked that the remaining contributions of $u\bar{u}$ resonances are negligible. There are, however, other resonance contributions that may be relevant in our case: the ones where the photon is attached to the initial and final state mesons in WA contributions (Fig. 1c-1d).

 In the spirit of the phenomenological Lagrangians, we would like to 
focus in the $q^2<2$ GeV$^2$ region by adding the effect of the lowest-lying light-flavoured resonances and pseudoscalar mesons as active degrees of freedom as it is done in the (Resonance) Chiral Lagrangians. This is the approach that we follow to ensure the applicability of the QCDf analysis of the 
$B^- \to P^- \ell^+ \ell^-$ decays down to threshold. 
Here we treat the photon-meson interaction by considering the exchange of resonances in the framework of two different models which have shown to give a very accurate description of experimental data for $q^2$ values up to $2$ GeV$^2$. This agreement shows that up to those energies, the photon does not resolve the quark structure of mesons and only probes their electromagnetic structure. In order to avoid potential double counting of effects from WA contributions (in the QCDf approach and ours), we will consider our LD WA contribution only for photons with $q^2 \lesssim 2$ GeV$^2$ and match it to the QCDf contribution for $q^2\gtrsim 2$ GeV$^2$ as discussed in section \ref{matching}.

The purpose of this paper is to quantify these expectations and evaluate the contributions given by diagrams in Figures 1(c,d) for 
the $B^- \to P^-\ell^+\ell^-$ decays for low and intermediate values of the invariant mass of di-lepton pair (other diagrams present in figure 2 vanish due to gauge invariance, see below). We use two different approaches for the $\pi^-$ and $K^-$ 
electromagnetic form factors in order to estimate the systematic error of our computation. For these rare $B$ decays, it is found that the one-photon exchange 
diagrams do not modify lepton universality for the energy range measured by LHCb \cite{lhcb} and that they turn out to be very small in the $P=K$ case, although their 
effects in the $P=\pi$ rates are more significant.

The WA contribution that we study has a different dependence on CKM mixing elements than the dominant SD one and also carries a sizable strong phase, leading to possible CP violating effects. The dominant source of CP violation, both in our paper and in the QCDf approach, arises from the interference between the SD top-mediated loop and the WA diagrams. However, the strong phase in our case is provided by the imaginary parts of the light resonance shapes, while in the QCDf approach it arises from the on-shell spectator $u$ quark in the $B$ meson after having emitted the photon. We find measurable effects both for $P=\pi,\,K$ on the integrated CP asymmetry and encourage the LHCb and Belle-II Collaborations to measure such asymmetries which could be useful to validate the models employed in the description of exclusive $b\to (s,d)\ell^+\ell^-$ decays at low photon virtualities.
\begin{figure}[!t]
\centering
\includegraphics[scale=0.7]{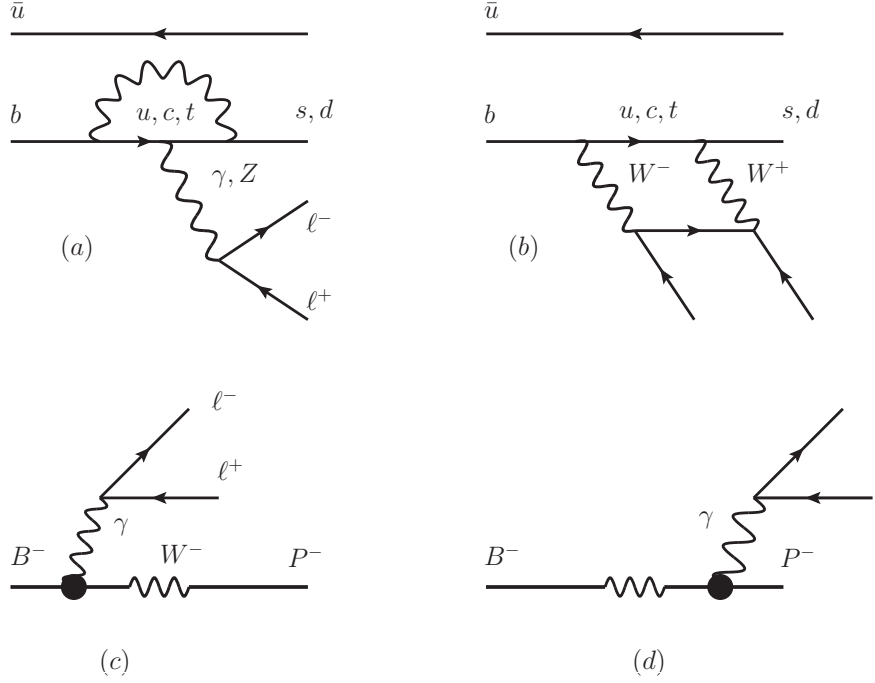}\
\caption{\small Short-distance contributions to $B^- \to P^- \ell^+\ell^-$ decays are shown in $(a)$ (penguin) and $(b)$ ($W$ box)
 diagrams. The one-photon, LD, WA contributions are shown in $(c)$ and $(d)$. The full circle denotes the electromagnetic
 form factor of the charged pseudoscalar mesons. Other long-distance structure-dependent vector and axial-vector terms (shown in fig. 2) do not contribute due to gauge invariance \cite{ecker}.}
\label{Fig:Leptonpair}
\end{figure}

\section{Decay amplitude}
In the Standard Model, observables like direct CP asymmetry in $B^{\pm}\to P^{\pm}\ell^+\ell^-$ have shown to be particularly sensitive to the interference of top dominated loops and WA amplitudes, which carry different weak and strong phases. Both amplitudes have been evaluated within the framework of QCDf where the strong phase in the WA amplitude, at relatively low photon virtualities, arises mainly from the on-shell $u$ spectator quark in the $B$ meson after having emitted the photon. In this Section we provide an alternative evaluation of this WA amplitude using the Resonance Chiral Theory within a factorization approximation, which we expect to be valid for $q^2\leq 2$ GeV$^2$. In a later Section we will show how to match this contribution to the corresponding WA amplitude calculated within QCDf, which will be used for larger $q^2$ values. Since the strong phase in our calculation stems from the pseudoscalar electromagnetic form factors, our approach provides an independent evaluation of the CP asymmetry for low and intermediate values of the lepton-pair invariant masses. 

\subsection{QCDf description}
For completeness and later use, we will consider first the 
description of these decays within QCDf as given in ref.~\cite{bobeth}. Let us choose
 the following convention for the particle's momenta: 
\[ B^-(p_B)\to P^-(p_P)\ell^+(p_+)\ell^-(p_-) \ . \] 
The  
contributions of the diagrams in Figures (1a,b), including higher order QCD corrections, to the decay amplitude can be written, to a very good approximation as \cite{bobeth}:
\begin{equation}\label{eqQCDf}
{\cal M}_{QCDf}=\frac{G_F\alpha}{\sqrt{2}\pi} V_{tb}V^*_{tD}\xi_P(q^2)p_B^{\mu}\left[ F_V(q^2)\bar{\ell}\gamma_{\mu}\ell+
F_A(q^2) \bar{\ell}\gamma_{\mu}\gamma_5 \ell \right]\ ,
\end{equation}
where $q=p_++p_-$ is the total momentum of the lepton pair, the subindex in the CKM matrix element stands for $D=d,s$ in the case
 of $|\Delta S|=0,1$ transitions. The factors $F_{V}\approx C_9=4.214$ and $F_A=C_{10}=-4.312$ denote the vector and axial Wilson
 coefficients at NNLO \cite{beneke} corresponding to the $O^q_{9}=(\bar{q}_L\gamma_{\mu}b)(\bar{\ell}\gamma^{\mu}\ell)$ and 
 $O^q_{10}=(\bar{q}_L\gamma_{\mu}b)(\bar{\ell}\gamma^{\mu}\gamma_5\ell)$ operators in the effective weak Hamiltonian for the $b\to q\ell^+\ell^-$ transition. The WA contributions induced by four quark operators are conventionally absorbed into the effective $F_V$ coefficient, by replacing its numerical value given above by $\xi_P(q^2)F_V\to \xi_P(q^2) F_V+F^{WA}$.
 For the $P=\pi$ case, the product of CKM matrix elements for loops involving all up-type quarks ($V_{ib}V_{id}^*$) scale as $\mathcal{O}(\lambda^3)$. However, the loop function favors the top quark as the 
 dominant contribution~\footnote{We are not neglecting the charm and up quark contributions in our numerical analysis, however. We are also including subleading 
 corrections with heavy-meson form factor ratios and other Wilson coefficients to eq.~(\ref{eqQCDf}), see ref.~\cite{bobeth}.} owing to the hierarchy of up-type quarks.

 Although we are aware that more refined heavy-to-light meson form factors have been developed \cite{Khod, otherFFs}, we will use
 the classical Ball and Zwicky form factors \cite{BZ} for our estimates and include the corresponding uncertainty in the errors, as 
discussed below. The expressions for the $q^2$-dependent form factors are 
\begin{eqnarray}
\xi_{\pi}(q^2)&=&\frac{0.918}{1-q^2/(5.32\ {\rm GeV})^2}-\frac{0.675}{1-q^2/(6.18\ {\rm GeV})^2}+{\cal P}_{\pi}(q^2), \nonumber \\
\xi_{K}(q^2)&=&\frac{0.0541}{1-q^2/(5.41\ {\rm GeV})^2}+\frac{0.2166}{\left[1-q^2/( 5.41\ {\rm GeV})^2\right]^2} +{\cal P}_K(q^2)\nonumber ,
\end{eqnarray}
where the polynomials ${\cal P}_P(q^2)$ as well as the needed Gegenbauer moments at the required energy scale can be found in 
Refs.~\cite{BZ, bobeth}.

\subsection{LD weak annihilation within R$\chi$T}

In this section, we would like to consider specifically light-resonance effects in the WA contribution to $B^\pm\to (\pi/K)^\pm \ell^+\ell^-$ for the $q^2\lesssim 2$ GeV$^2$ region. We assume that in this energy range the relevant degrees of freedom are these light mesons and resonances (and not the quark and gluon fields), according to the methodology of phenomenological Lagrangians. Low-energy phenomenology supports widely this hypothesis \cite{EFTs}. Applying this method to the process under consideration requires further hypothesis about  the strong and electromagnetic interactions of $B$ mesons which, we argue, play a subleading role in our description.

The decay amplitude corresponding to WA is given by

\begin{equation}
\mathcal{M}_{LD}^{WA}=\frac{e^2}{q^2}\bar{\ell}\gamma^{\mu}\ell\, \mathcal{M}_{\mu}^{WA} 
\end{equation}
where $\mathcal{M}_{\mu}^{WA} $ denotes the effective hadronic electromagnetic current coupled to the leptonic current. Conservation of the electromagnetic current demands 
\begin{equation}
 \mathcal{M}_{\mu}^{WA}=\left[(p_B+p_P)_\mu-\frac{m_B^2-m_P^2}{q^2}q_\mu\right]F(q^2) \ ,
\end{equation}
where only the first term within square brackets gives a non-vanishing contribution owing to the condition $q^{\mu} \bar{\ell}\gamma_{\mu}\ell=0$ (for the same reason, we can also replace $(p_B+p_P)_\mu\to 2p_{B\mu}$). The factor $F(q^2)$ encodes the information about the dynamics of weak and electromagnetic interactions in the hadronic blob.

\begin{figure}[!h]
\centering
\includegraphics[scale=0.7]{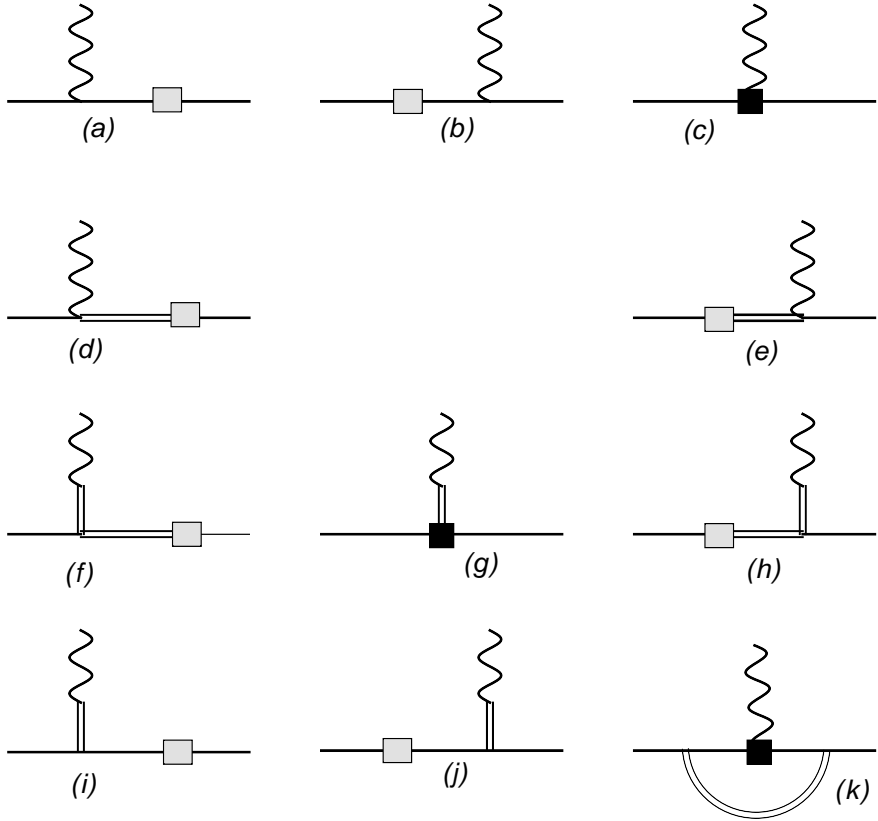}\
\caption{\small Some of the Feynman diagrams contributing to the effective hadronic electromagnetic  $B^{\pm} \to P^{\pm}\gamma^*$ vertex. Single lines stand for pseudoscalar mesons, double lines for (axial-)vector resonances and wavy lines for the virtual photon (due to the spin-one nature of the weak current, spin-zero resonance contributions are suppressed). Filled squares denote the weak/electromagnetic vertex, while the empty rectangles denote the WA Hamiltonian.  All contributions to the $B^\pm\to P^\pm \ell^+ \ell^-$ decays (including the pointlike interactions in the first line) vanish due to gauge invariance \cite{ecker} or are suppressed, except diagrams (i) and (j) which contribute to the electromagnetic form factors of pseudoscalar mesons.  Odd-intrinsic parity violating vertices are also considered in the calculation.}
\label{newfig}
\end{figure}

Owing to gauge-invariance, most of the diagrams (shown in Figure \ref{newfig}) contributing to WA, vanish. To illustrate this, we insert a complete set of intermediate states in the relevant hadron matrix element and take into account the T-ordering of the electromagnetic and weak interactions, which gives rise to the leading order factorizable contributions
\begin{eqnarray}
\mathcal{M}_{\mu}^{WA}&=&\int d^4x e^{-iq\cdot x} \left\langle P^{\pm} |T\left\lbrace j^{\rm em}_{\mu}(x) \mathcal{H}^{WA}_{\rm eff}(0)\right\rbrace|B^{\pm}\right\rangle\nonumber\\
&=& \left\langle P^{\pm}|j^{\rm em}_{\mu}|X^\pm\right\rangle \left\langle X^{\pm}|\mathcal{H}^{WA}_{\rm eff}|B^\pm\right\rangle \nonumber\\
&&\ + \left\langle P^{\pm}|\mathcal{H}^{WA}_{\rm eff}|Y^\pm \right\rangle \left\langle Y^{\pm}|j^{\rm em}_{\mu}|B^\pm\right\rangle \ ,
\end{eqnarray}
 where the sum over virtual intermediate states $X$ and $Y$ that are allowed by the quantum numbers of the weak and electromagnetic interactions, should be understood.

 Some of the Feynman diagrams that contribute to $\mathcal{M}_{\mu}^{WA}$ are depicted in Fig. \ref{newfig}. Non-factorizable contributions, like the ones shown in Figure \ref{newfig}k, are sub-leading in the $1/N_C$ expansion and will not be considered in our calculation \footnote{We recall that R$\chi$T is an expansion in $1/N_C$} . At lower photon virtualities, where resonance degrees of freedom may be neglected, it was shown that non-factorizable (local) contributions partially cancel among themselves\cite{ecker}, leaving the form factors of $K$ and $\pi$ as the dominant contributions in $K^{\pm}\to \pi^{\pm}\ell^+\ell^-$.

Under the above approximations, which justify the naive factorization of the hadronic matrix element of the weak hamiltonian, the leading order one-photon exchange (WA) amplitude corresponding to Figure~\ref{newfig}(i,j) 
can be computed taking into account that
\begin{equation}
\left\langle0|\bar{u}\gamma^\mu(1-\gamma_5)b|B^-\right\rangle=-i\,f_B\,p_B^\mu\,,\quad \left\langle P^-|\bar{D}\gamma_\mu(1- \gamma_5) u|0\right\rangle=i\,f_P\,p_{P\,\mu}\,,
\end{equation}\label{derivativecouplings}
and is given by (neglecting the contribution of diagram 1c --which is of order $(m_P/m_B)^2$ with respect to 1d-- 
  is, however, a good approximation \footnote{The radiation emitted from lighter charged particles is expected to dominate for low values of photon virtualities in agreement with our result in eq.~(\ref{eqLD}).})
\begin{eqnarray}\label{eqLD}
{\cal M}_{LD,WA}&=&\sqrt{2}G_F(4\pi\alpha)V_{ub}V_{uD}^*f_Bf_P\frac{1}{q^2(m_B^2-m_P^2)} \nonumber \\
&& \ \times \left[M_B^2 \left(F_P(q^2)-1\right)- 
m_P^2\left(F_B(q^2)-1\right)\right] p_B^{\mu}\bar{\ell}\gamma_{\mu}\ell\,,
\end{eqnarray}
where $f_X$ denotes the decay constant of the pseudoscalar meson $X$ according to the PDG \cite{pdg2014} conventions for 
$f_{K, \pi, B}$ and $F_{X}(q^2)$ is the electromagnetic form factor of the corresponding meson. Note that the pure scalar QED 
(point meson approximation), as well as the structure-dependent vector and axial-vector terms, do not
 contribute to the amplitude as required by gauge invariance \cite{ecker} \footnote{In particular, let us stress that this forbids contributions from $\pi\rho\gamma$ or $\pi a_1\gamma$ vertices for the $B^- \to \pi^-\ell^+\ell^-$ transition (and analogous vertices for the $K$ case).}. Thus, the one-photon exchange contribution is sensitive to the intermediate $q^2$ region of the pseudoscalar form factor, and eventually to their resonance structure.

  Due to the vector nature of the one-photon exchange contribution, its amplitude can be absorbed into the contribution of the $O_9$ operator in the QCDf 
amplitude 
  under the replacement 
\begin{equation}
\xi_P(q^2)F_V \longrightarrow \xi_P(q^2)F_V+\kappa_Pm_B^2\left[\frac{F_P(q^2)-1}{q^2}\right],
\end{equation}
where 
\begin{equation}
\kappa_P=-8\pi^2\frac{V_{ub}V_{uD}^*}{V_{tb}V_{tD}^*}\frac{f_Bf_P}{m_B^2-m_P^2}\,.
\end{equation}
Note that $\kappa_{P}\sim \mathcal{O}(10^{-2})\times \frac{V_{ub}V_{uD}^*}{V_{tb}V_{tD}^*}$ so that its influence is governed by the ratio of CKM factors which is 
$\sim \mathcal{O}(\lambda^0)$ for $P=\pi$ and $\mathcal{O}(\lambda^2)$ for $P=K$. This suggests a larger effect for $B^-\to \pi^-\ell^+\ell^-$ transitions but a 
detailed analysis of the electromagnetic meson form factors is needed to confirm these expectations.

\section{Pseudoscalar form factors}

 For the electromagnetic form factors of the light pseudoscalar mesons ($P=\pi,\,K$) we have considered two approaches. On the one 
hand we have used form factors that are obtained within the frame of Resonance Chiral Theory (R$\chi$T) \cite{RChT} and, on the other hand, 
the phenomenological form factors used by the BaBar Collaboration which employs the Gounaris-Sakurai (GS) parametrization \cite{GS}. The 
first approach has the advantage of providing a low-energy behaviour complying with the chiral limit of QCD \cite{ChPT}, which is a 
must if we want to get close to thresholds in some of our evaluations \footnote{On the contrary, GS parametrizations introduce 
spurious phases below thresholds and the decoupling of excited resonances at low energies is spoiled.}. Alternatively, the GS
 parametrizations include more excited resonances and, for this reason, are expected to give a closer description of data at higher
 energies. In the cases at hand, the bulk of the contribution (even for integrated observables starting at $q^2=1$ GeV$^2$) 
will be given by the $\phi(1020)$ ($\rho(770)$) meson exchange in the K ($\pi$) cases. Therefore, we should expect very similar results 
for the two approaches, $R\chi T$ being more reliable for observables starting near thresholds and the GS for the much less important 
higher energy range (up to $6$ or $8$ GeV$^2$ depending on the channel).

The minimal Lagrangian of $R\chi T$ was derived in Ref.~\cite{RChT} upon the requirement of chiral symmetry for the pseudoGoldstone fields and flavor symmetry for the light-flavored resonances. Ref. \cite{RChT} supplements the $\chi PT$ Lagrangian at lowest order ($p^2$) in the chiral counting \cite{ChPT} \footnote{The conventions used in the following are standard and can be checked, for instance, in \cite{Cirigliano:2006hb}.}

\begin{equation} \label{eq:op2}
{\cal L}_{\chi {\rm PT}}^{(2)}=\frac{F^2}{4}\langle u_{\mu} u^{\mu} + \chi _+ \rangle \ ,
\end{equation}
with the following interaction terms between a light spin-one resonance and the pseudoGoldstone fields \cite{RChT}

\begin{eqnarray} \label{eq:lagRChT}
{\cal L}_2^{\mbox{\tiny V}} & = &  \frac{F_V}{2\sqrt{2}} \langle V_{\mu\nu}
f_+^{\mu\nu}\rangle + i\,\frac{G_V}{\sqrt{2}} \langle V_{\mu\nu} u^\mu
u^\nu\rangle  \; \, , \nonumber \\
{\cal L}_2^{\mbox{\tiny A}} & = &  \frac{F_A}{2\sqrt{2}} \langle A_{\mu\nu}
f_-^{\mu\nu}\rangle \;.
\end{eqnarray}

The real couplings $F_V$, $G_V$ and $F_A$ are unrestricted by symmetries and encode the dynamics. One can proceed similarly for the spin-zero resonances \cite{RChT} and include terms with more resonance fields and excited resonances, extending also to the odd-intrinsic parity sector \cite{Cirigliano:2006hb, extensionsRChT}. Within the single resonance approximation, the evaluation of the $P$-vector form factor using the $R\chi T$ Lagrangian yields

\begin{equation}
F_P(q^2)\,=\,1+\frac{F_V G_V}{F^2}\frac{q^2}{M_V^2-q^2}\,.
\end{equation}

By demanding a Brodsky-Lepage behaviour \cite{BL} to $F_P(q^2)$ one gets the relation $F_V G_V = F^2$, which gives a parameter-free expression. This naive result can be improved by including the next-to-leading order effect in the $1/N_C$ counting, {\it i. e.} the meson (off-shell) width according to chiral constraints (the corresponding real part of the chiral
loops is also included), and adding the effect of higher excitations and (in the $\pi$ case) including the dominant isospin-breaking effect given by the $\rho-\omega$ mixing.

 Specifically for $F_\pi(q^2)$, we have used the parametrizations in the last two Refs. in \cite{pionff} \footnote{The numerical values that we have employed in 
 the present analysis are those given in these references.}, which include three isovector resonances 
($\rho(770)$, $\rho(1450)$ and $\rho(1700)$) and the resummation of final state interactions encoded in the chiral loop functions. These representations provide 
good quality fits to Belle data \cite{Belle2pi} for $\tau^-\to\pi^-\pi^0\nu_\tau$ decays. Additionally, we have included the characteristic $\rho(770)-\omega(782)$ 
interference appearing in the neutral channel by multiplying the $\rho(770)$ term by the factor 
\begin{equation}
 1-\theta_{\rho\omega}\frac{q^2}{3M_\rho^2}\frac{1}{M_\omega^2-q^2-iM_\omega\Gamma_\omega}\,,
\label{rhoomegamixing}
\end{equation}
with $\theta_{\rho\omega}=(-3.3\pm0.5)\cdot10^{-3}$ GeV$^2$ \cite{Pich:2002ne}. Other isospin breaking corrections are neglected.

This parametrization is compared to BABAR data \cite{pionff-babar} in Figure 2, which are available for the energy region $2m_{\pi} \leq 
\sqrt{q^2} \leq 3$ GeV. There, the phenomenological GS parametrization includes an additional $\rho$-like excitation. 
Tiny differences between both parametrizations are seen in the region where the destructive interference between the $\rho(1450)$ 
and the $\rho(1700)$ resonances is stronger. The comparison also shows the effect of the $R\chi T$ parametrization lacking of the 
$\rho(2250)$ meson, which is clearly visible in the $(2,2.3)$ GeV region. These minor differences are taken into account in the 
final quoted errors.

\begin{figure}[!t]
\centering
\includegraphics[scale=0.6,angle=-90]{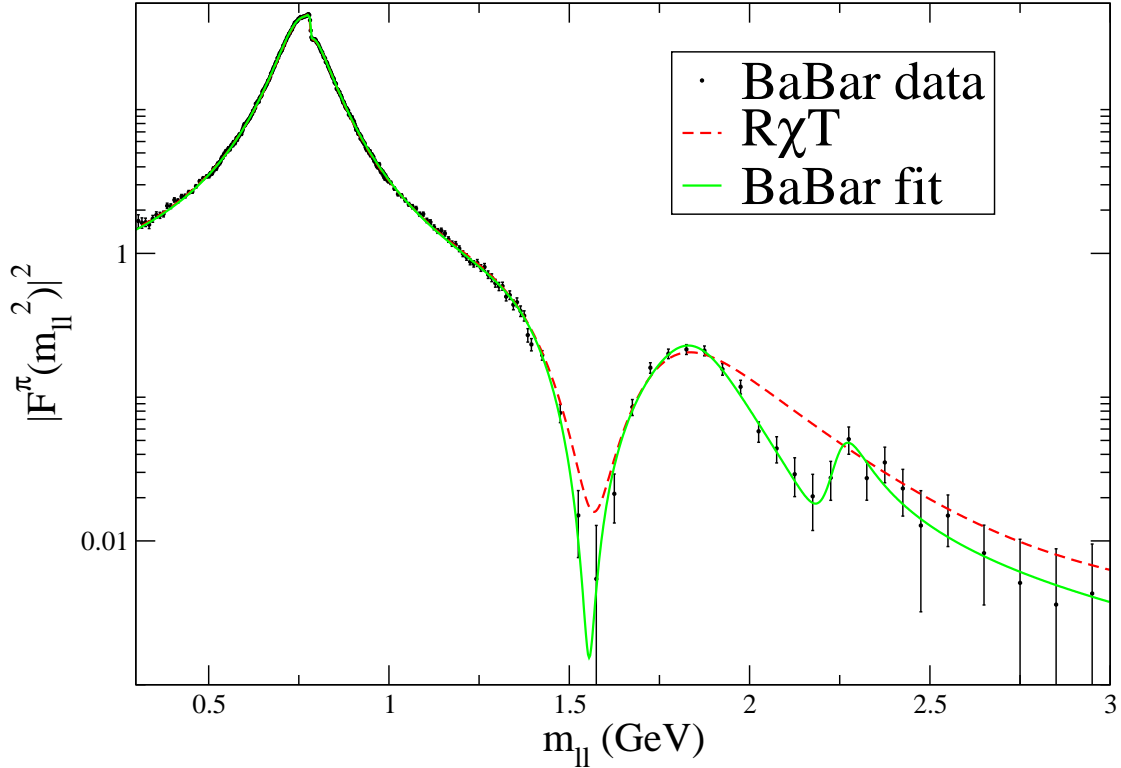}\
\caption{\small R$\chi$T and GS parametrization (BaBar fit) of the electromagnetic pion form factor as a function of $m_{ll}= \sqrt{q^2}$ are compared to 
experimental data from BABAR \cite{pionff-babar}.}
\label{pionfff}
\end{figure}

 Analogously, we have employed two parametrizations of $F_K(q^2)$ form factor. Since this form factor is completely dominated by the extremely 
narrow $\phi(1020)$ meson, we have considered the $R\chi T$ prediction with only one multiplet of resonances \cite{kaonff}.
 On the other hand, the BaBar Collaboration \cite{kaonff-babar} has reported measurements of the kaon form factor from threshold up
 to $2.5$ GeV. Since $|F_K(q^2)|^2$ drops by 6 orders of magnitude in going from the peak to $2.5$ GeV 
(see Figure 3), it should be sufficient to include only the $\phi(1020)$ resonance in its parametrization.

 As it can be observed in Figure 3, experimental data for $|F_K(q^2)|^2$ are reasonably well described from threshold up to approximately $1.3$ GeV. 
 Deviations at higher energies have a negligible impact in the integrated observables of rare $B$ decays. Above this energy, other resonance structures
 with very small (and alternating) peaks and dips around the single resonance queue can be observed. These, in turn, are well 
described using the parameterization quoted in the BaBar paper \cite{kaonff-babar} (which we fitted to the BaBar data), 
which includes two $\phi$, three $\rho$ and three $\omega$ excitations in addition to the single lightest vector meson multiplet included in the $R\chi T$ form
factor \footnote{Besides, in the limit of ideal $\omega(782)-\phi(1020)$ mixing the former state does not contribute to $F_K(q^2)$ and only $\rho(770)$ and 
$\phi(1020)$ remain, with a prominent role of the latter.}.

\begin{figure}[!t]
\centering
\includegraphics[scale=0.6, angle=-90]{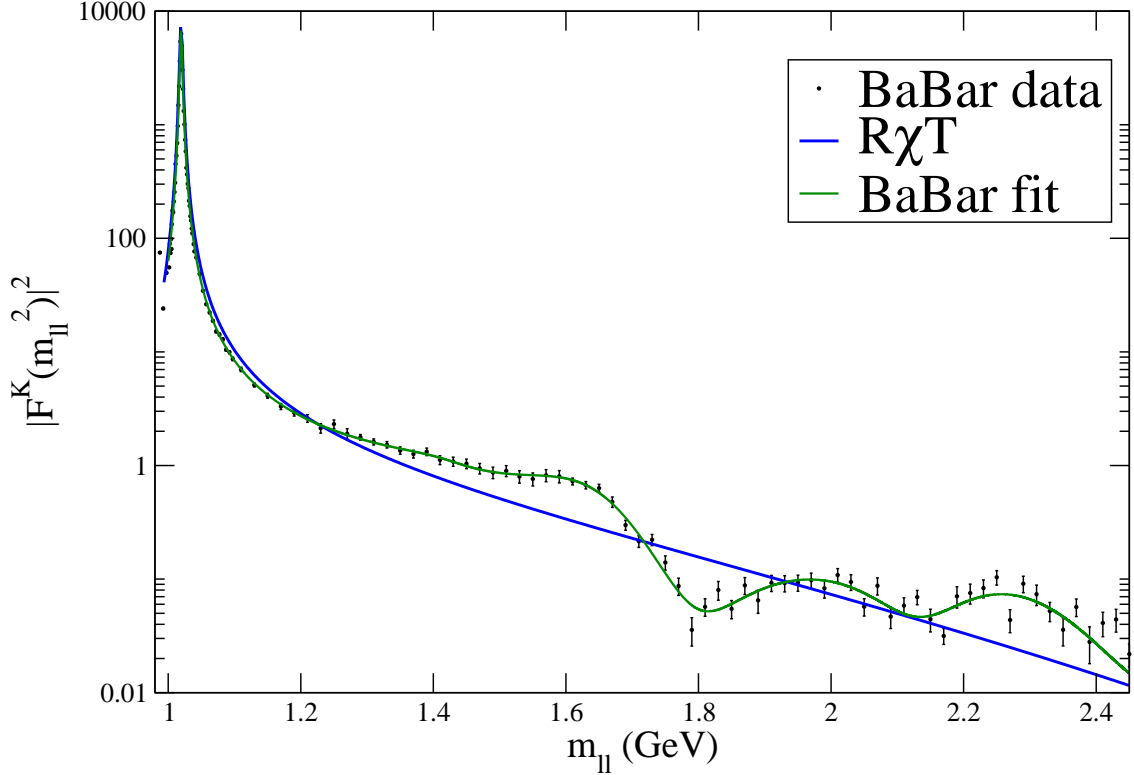}\
\caption{\small R$\chi$T and GS parametrization (BaBar fit) of the electromagnetic kaon form factor as a function of $m_{ll}=\sqrt{q^2}$ are compared to 
experimental data from BABAR \cite{kaonff-babar}.}
\label{Fig:Leptonpair}
\end{figure}

Because of its relevance in the study of CP violating observables, we plot in Fig. \ref{ReImparts} the real and imaginary parts of the light-meson electromagnetic form factors according to the two different descriptions that we used. While models  
agree nicely for the $\pi$ case, they do not in the $K$ case around the $\rho(770)$ peak. Since the corresponding phase shift has been validated by data in the 
case of the $R\chi T$ description, we attribute this to an incorrect phase of the $GS$ kaon form factor.

\begin{figure*}[t]
\centering
\includegraphics[scale=0.4, angle=-90]{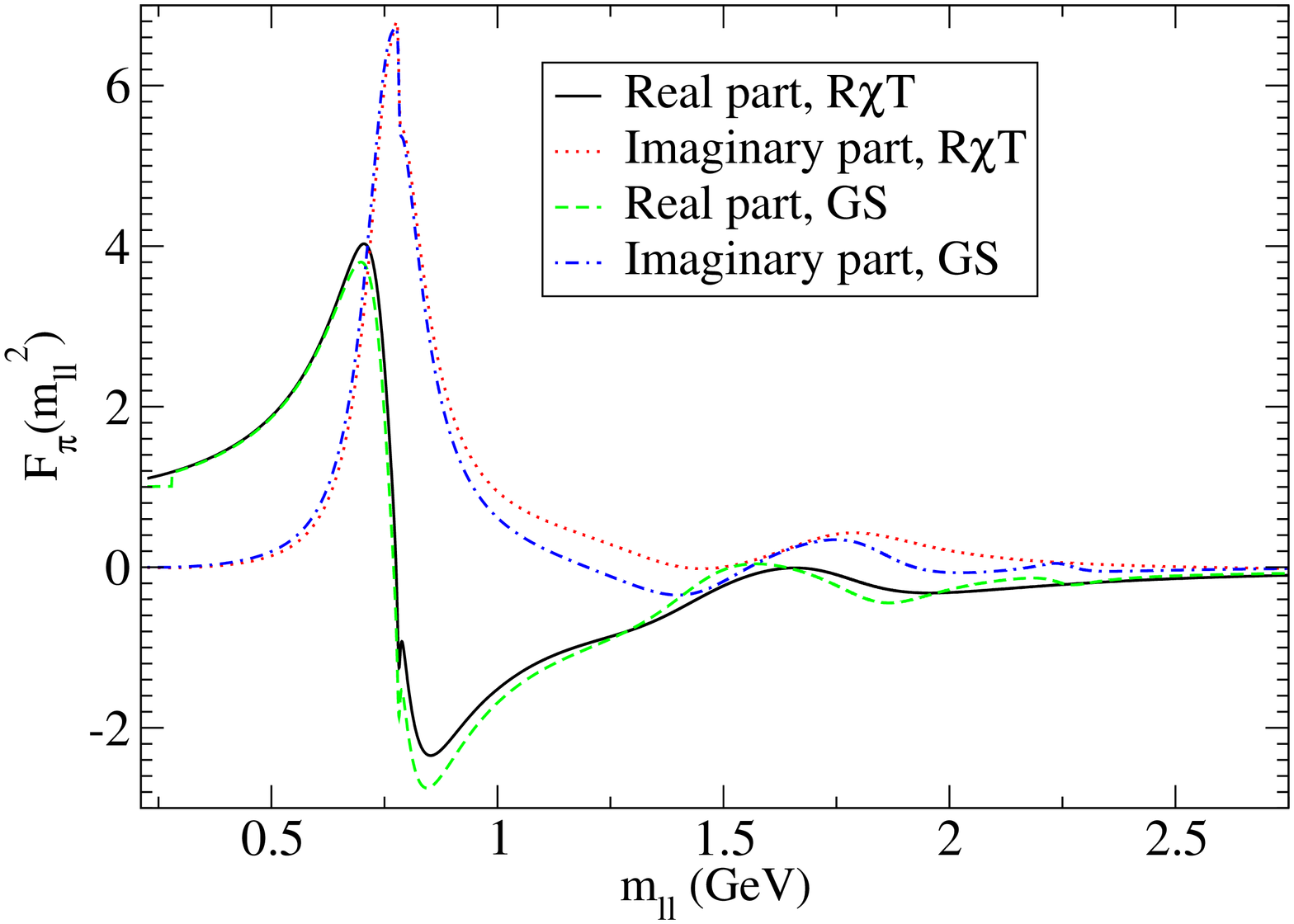}
\end{figure*}
\vspace*{1.0cm}
\begin{figure}[t]
\centering
\includegraphics[scale=0.4, angle=-90]{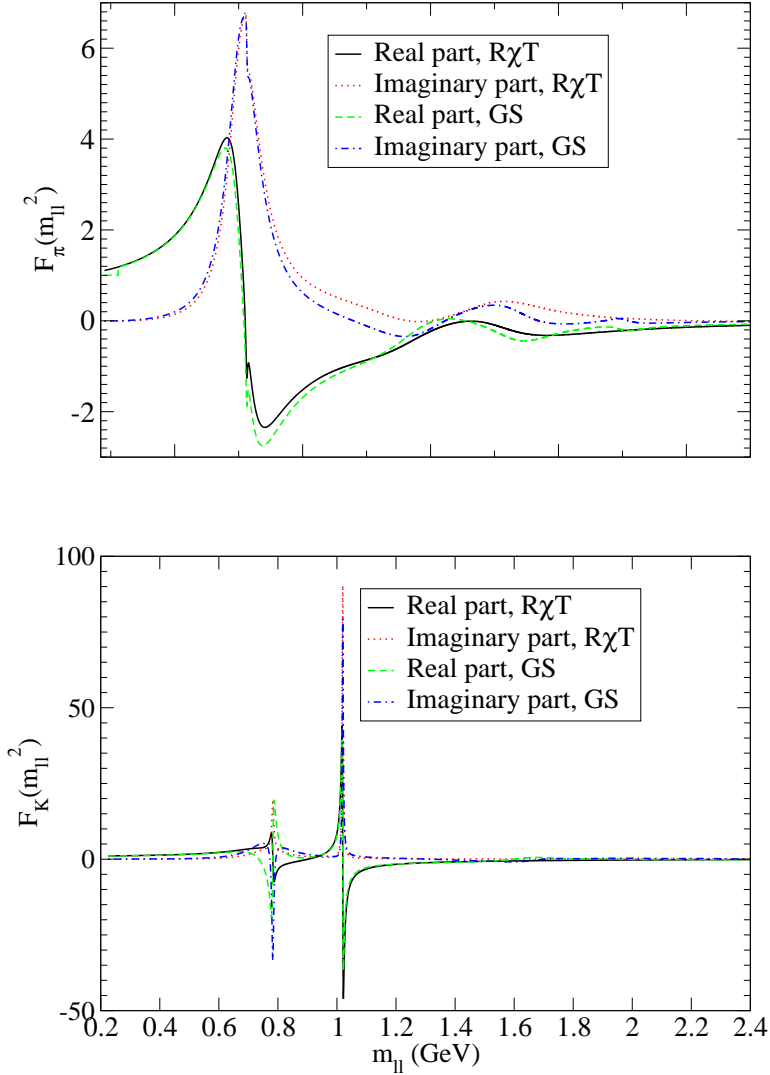}
\caption{\small Real and imaginary parts of the electromagnetic light-meson form factors for $P=\pi$ (upper half) and $P=K$ (lower half) as a function of $m_{\ell\ell}=\sqrt{q^2}$, according to the $R\chi T$ and $GS$ descriptions.}
\label{ReImparts}
\end{figure}

 Finally, the $B$-meson electromagnetic form factor contribution is suppressed by a factor $m_P^2/M_B^2$ in eq.~(\ref{eqLD}) with respect to the ones of lighter 
 mesons. In addition, a dynamical suppression of the form factor itself is expected according to the reduced charged radius of heavy mesons. 
Just in order to confirm this  \footnote{The only way to access the electromagnetic $B$ meson form factor at small values of the momentum transfer, would be by means of elastic electron-$B$ meson scattering. Given the difficulty to produce a $B$ meson target/beam, it becomes rather complicated to test this low energy behavior of the  $B$ meson form factor.}, we consider the effective Lagrangian coupling $D$-mesons to lighter mesons and external sources in 
ref.~\cite{Ds} generalized to include $B$ mesons\footnote{The procedure employed is consistent with the heavy quark mass limit, as 
explained in the quoted reference.}. It can be seen that, upon requiring a Brodsky-Lepage behaviour \cite{BL} of this form factor at
 infinite momentum transfer, the shape of $F_B(q^2)$ is given just by flavour symmetry, yielding
\begin{equation}
F_B(q^2)\,=\,1+\frac{3}{2}q^2\left(\frac{1}{M_\rho^2-q^2-i M_\rho \Gamma_\rho(q^2)}-\frac{1}{3(M_\omega^2-q^2-i M_\omega \Gamma_\omega)}\right)\,,
\end{equation}
where $\Gamma_\rho(q^2)$ can be found in Refs.~\cite{pionff}. Our numerical evaluations confirm that the effect of this form factor will be 
completely negligible in rates of rare $B^-$ mesons.

The spectra of the lepton pair, normalized to the $B^-$ meson decay width are plotted in Figure \ref{Fig:Leptonpair}  for low values of the lepton-pair
 invariant mass. We observe that above the muon-pair threshold both spectra are identical. Clearly, the integrated rates in the whole kinematical range would 
 exhibit a trivial breaking of leptonic universality owing to the lower electronic threshold and its enhancement due to the $1/q^2$ dependence in the LD WA amplitude. However, this 
 will happen both in the well-known QCDf as well as in the new LD WA contributions; thus only the numerical evaluation will tell if there is 
 any additional measurable breaking of universality due to LD WA effects in the considered processes.

\begin{figure*}[t]
\centering
\includegraphics[scale=0.4, angle=-90]{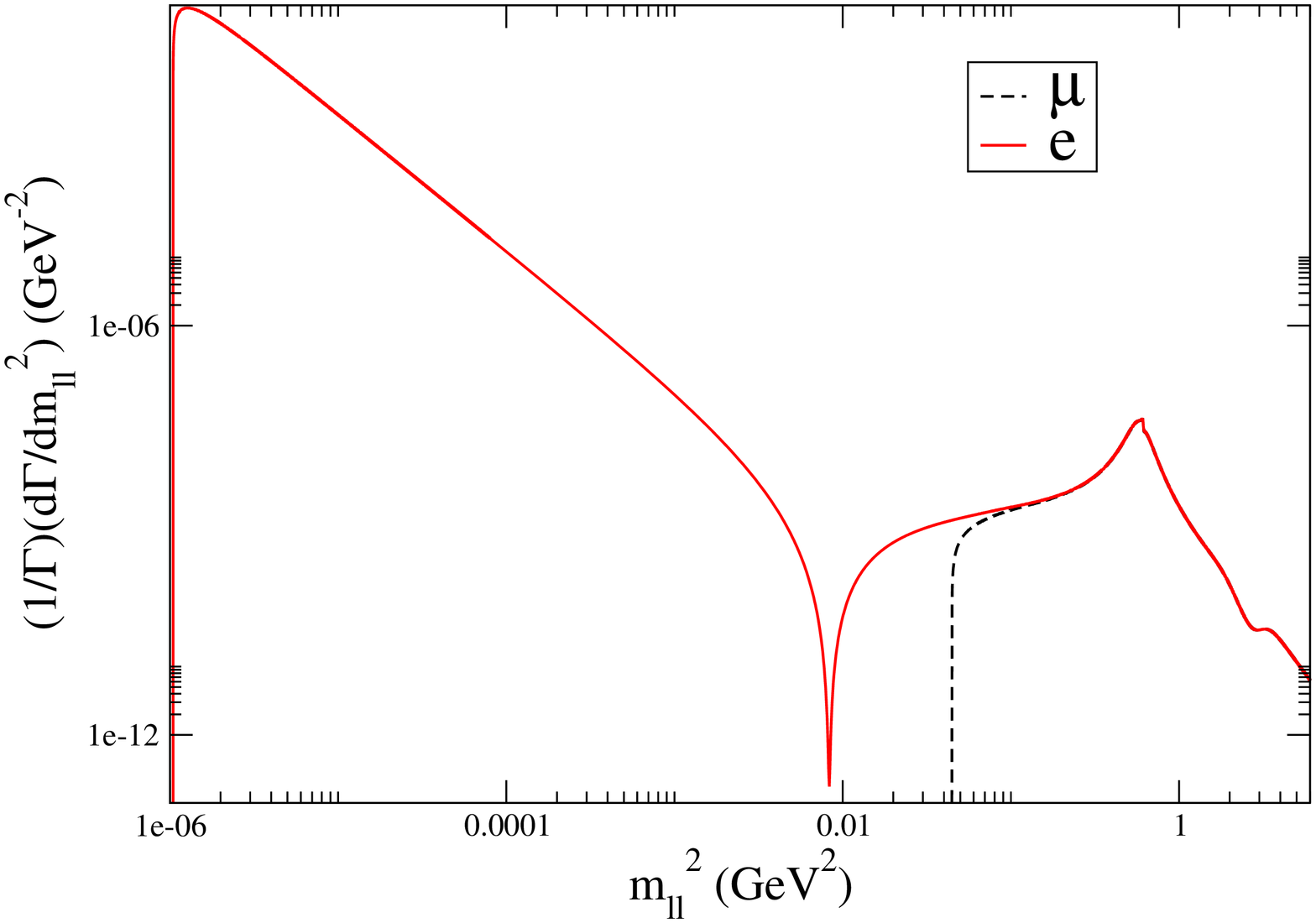}
\end{figure*}
\vspace*{1.0cm}
\begin{figure}[t]
\centering
\includegraphics[scale=0.4, angle=-90]{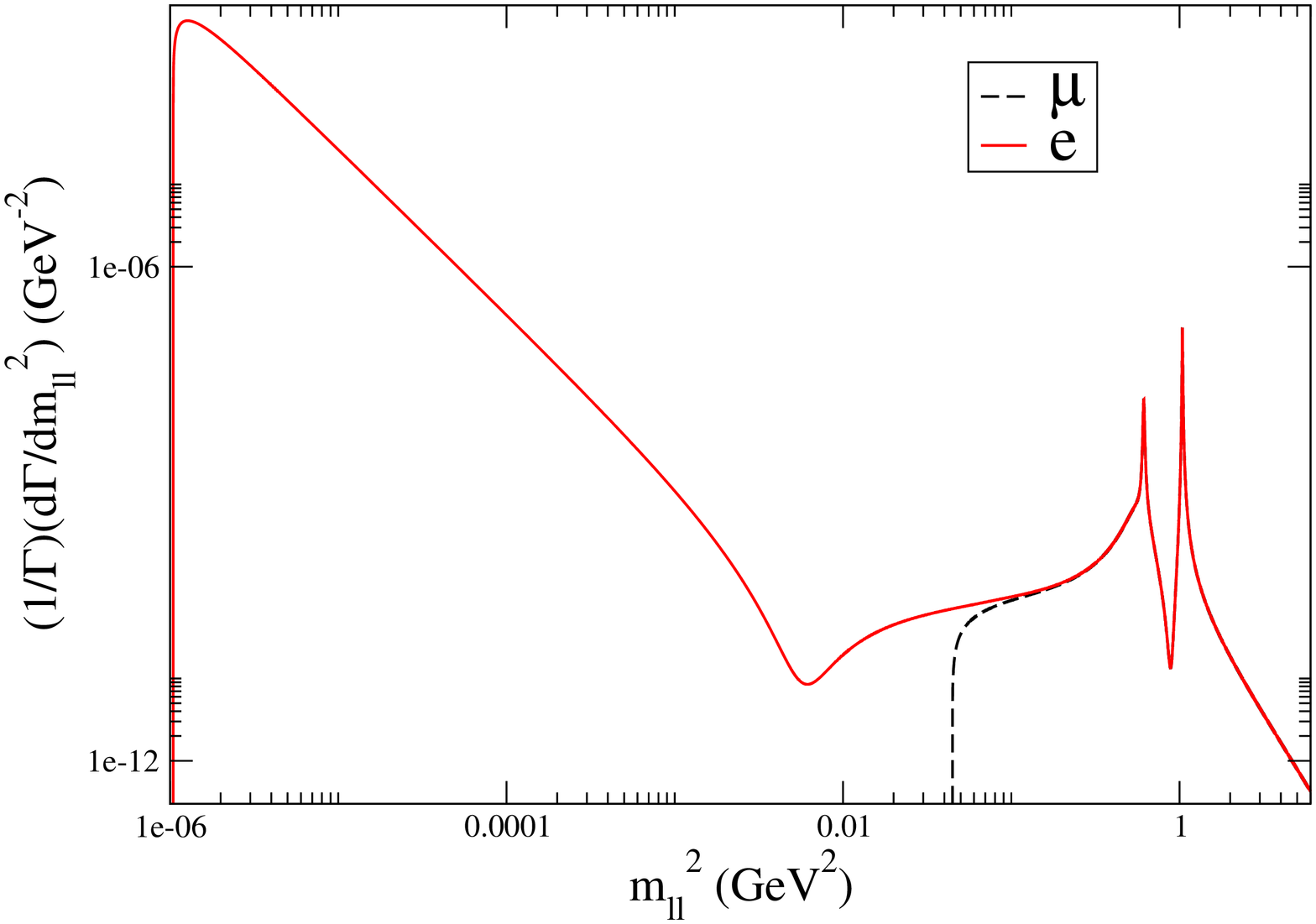}
\caption{\small Normalized lepton-pair spectrum for the LD WA contribution to $B^- \to P^-\ell^+\ell^-$ decays for $P=\pi$ (upper half) and $P=K$ (lower half). The
  squared lepton-pair invariant mass is taken from threshold up to $1.5$ GeV$^2$.}
\label{Fig:Leptonpair}
\end{figure}

\section{Matching of the R$\chi$T and QCDf descriptions of the WA contributions}\label{matching}

The LD (one-photon exchange) contribution to the $B^- \to P^-\ell^+\ell^-$ decays that we have discussed in previous sections and the one evaluated in the QCDf approach originate from the WA four-quark operator of the effective Hamiltonian. Then it naturally arises the question whether there is some double-counting between both contributions or not.

For low enough values of $q^2$ ($q^2\ll M_B^2$), the QCDf expansion in 1/$E_P$ is the best way to organize the perturbative series. However, when approaching the $u\bar{u}$ threshold ($q^2_{max}<M_{J/\Psi}$) light-resonance effects need to be included explicitely as active degrees of freedom in the action as done in the resonance chiral Lagrangians. Within the interval from threshold to $q^2_{max}$, we expect that chiral Lagrangians give an adequate description for low $q^2$ values, while QCDf be more appropriate for the higher $q^2$ region. We thus assume there is an intermediate energy scale where both descriptions are good approximations and search for this matching scale where the (complex) effective Wilson coefficient $F_V\equiv C_9^{eff}$ smoothly matches both descriptions for the $\pi$ and $K$ meson channels. We have found such matching scale at $q^2\sim 2$ GeV$^2$, as it can be seen in Figs.~\ref{Fig:Matching}.

Then we have consider our LD one-photon exchange diagrams as the WA contribution for $q^2<2$ GeV$^2$ and the SD QCDf counterpart (where the photon is radiated by the valence quarks in the $B$ and $P$ mesons) for $q^2>2$ GeV$^2$. We have used this separation scale between the LD and SD (QCDf) descriptions of WA contributions in the following and a slight variation of the matching scale has been taken into account in the error estimates. In the following, whenever we use QCDf it is understood that the corresponding result for WA is only used above $2$ GeV$^2$.

\begin{figure*}[t]
\centering
\includegraphics[scale=0.4, angle=-90]{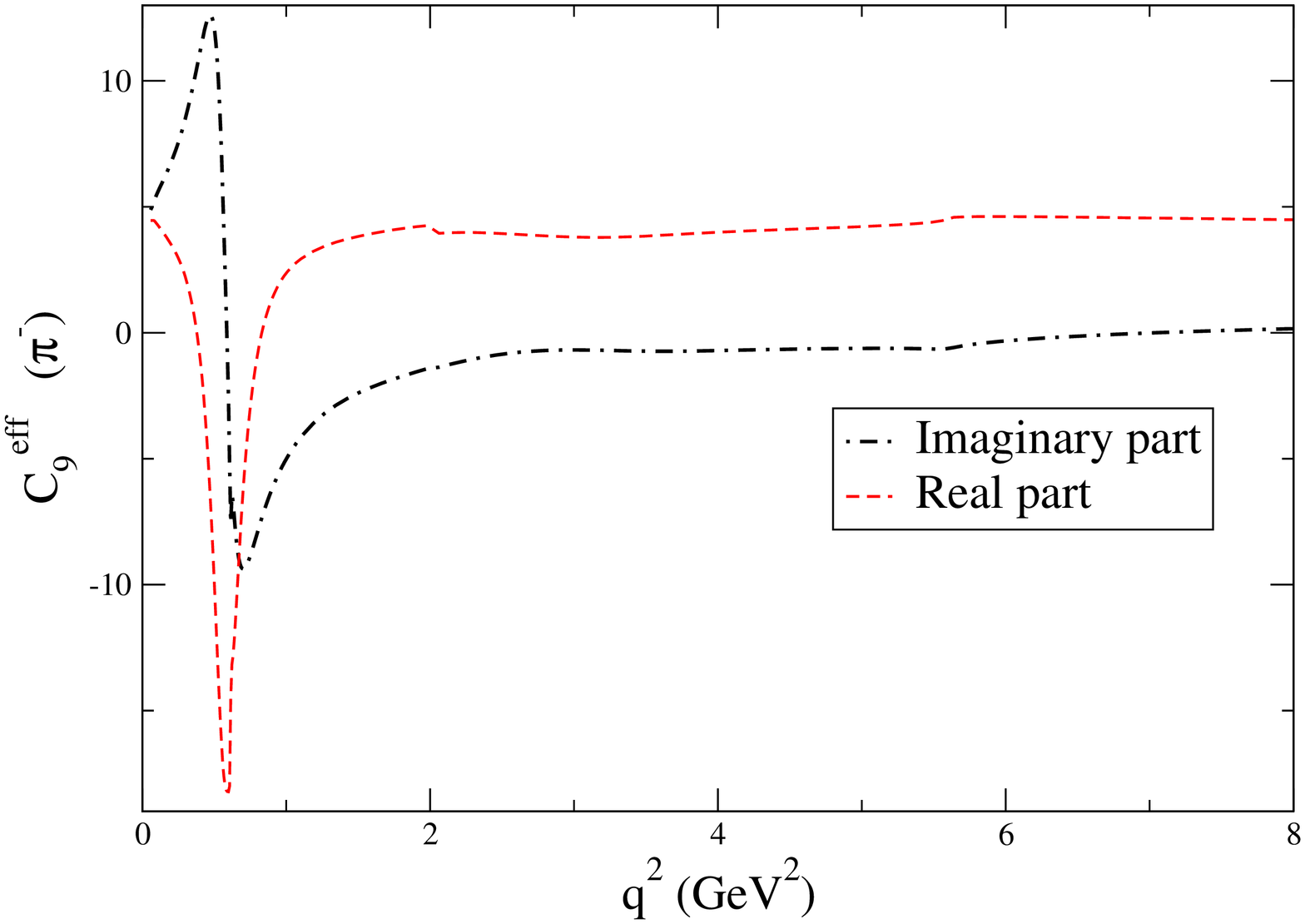}
\end{figure*}
\vspace*{1.0cm}
\begin{figure}[t]
\centering
\includegraphics[scale=0.4, angle=-90]{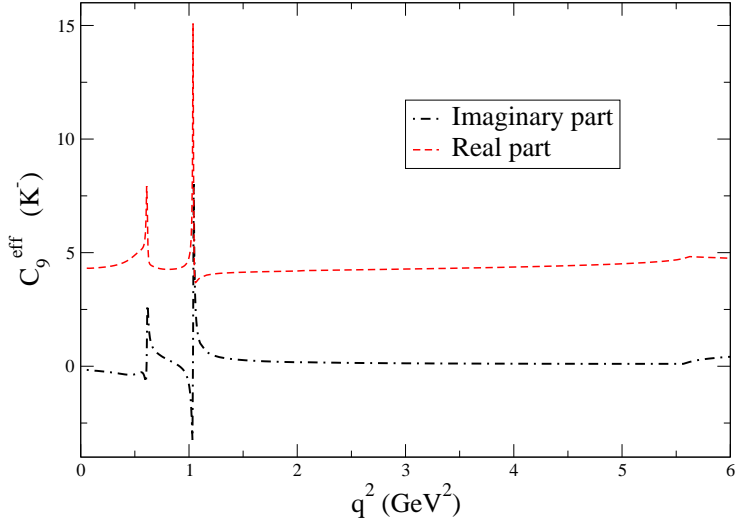}
\caption{\small A smooth matching between the LD and SD (QCDf) WA contributions is appreciated in the plot of $C_9^{eff}$ as a function of $q^2$ for $P=\pi$ (upper half) and $P=K$ (lower half).}
\label{Fig:Matching}
\end{figure}

\section{Branching fractions for $B^- \to P^-\ell^+\ell^-$}
Searches for New Physics at large hadronic recoil in the $B^- \to K^-\ell^+\ell^-$ decays are restricted to the $(1,\,6)$ GeV$^2$ 
range of $q^2$ \cite{lhcb}. Thus we will stick to this region for this channel. The $B^- \to \pi^-\ell^+\ell^-$ decays have just been observed \cite{lhcb2} and such studies have not taken place yet. We therefore include our results both for a range starting at 
$q^2=0.05$ GeV$^2$ (basically the muon threshold) and at $1$ GeV$^2$. In either case we cut the phase-space integration at $8$ GeV$^2$ to avoid the charmonium region. In Table \ref{Table1} we show the corresponding branching ratios of LD WA,  
QCDf 
and their interference contributions by considering these different kinematical integration domains. Values for required input parameters are taken from the PDG \cite{pdg2014} except for the CKM matrix elements which come from  Ref.~\cite{ckmfitter2015}.

\begin{table}[!t]
\caption{\small Integrated branching ratios of $B^-\to P^-\ell^+\ell^-$ decays for $P=\pi$ (left hand side) and $P=K$ (right hand side) for different $q^2$ ranges. We tabulate separately 
the 
QCDf
, long-distance WA (LD) and their interference contributions for the kinematical ranges of interest.}
\begin{tabular}{|p{3cm}|p{4cm}|p{4cm}|p{4cm}|}
\hline
\multicolumn{3}{|c|}
{$B^-\to \pi^-\ell^+\ell^-$}&{$B^-\to K^-\ell^+\ell^-$}\\
\hline
 & $0.05 \leq q^2 \leq 8$ GeV$^2$  & \ \ \ $1 \leq q^2 \leq 8$ GeV$^2$& \ \ \ $1 \leq q^2 \leq 6$ GeV$^2$ \\
 \hline
LD&$(9.06\pm0.15)\cdot 10^{-9}$& $(4.74\pm0.05)\cdot 10^{-10}$ & $(1.70\pm 0.21)\cdot 10^{-9}$\\
interf. &$(-2.57\pm 0.13)\cdot 10^{-9}$& $(-2^{+2}_{-1})\cdot 10^{-10}$ &$(-6\pm 2)\cdot 10^{-11}$\\
QCDf &$(9.57^{+1.45}_{-1.01})\cdot 10^{-9}$& $(8.43^{+1.31}_{-0.87})\cdot 10^{-9}$& $(1.90^{+0.69}_{-0.41})\times 10^{-7}$\\
Total & $(1.61^{+0.15}_{-0.11})\cdot 10^{-8}$ & $(8.69^{+1.31}_{-0.87})\cdot 10^{-9}$&$(1.92^{+0.69}_{-0.41})\times 10^{-7}$ \\
\hline
\end{tabular}\label{Table1}
\end{table}
 Several comments concerning these results are in order:
\begin{itemize}
 \item Our results for the 
QCDf 
contribution to the $B^-\to \pi^-\ell^+\ell^-$ branching ratio are higher than those in ref.~\cite{ali} because of the different 
 heavy-meson form factors employed. Specifically, in this analysis form factors parameters were fitted to reproduce $B^+\to \pi^0\ell^+\nu_\ell$ data, 
 resulting in smaller 
QCDf 
contributions than in other analyses \cite{btopi0, btopi} or ours.
 \item Another source of difference in the 
QCDf
 contributions comes from the updated inputs we are using \cite{ckmfitter2015}, while older PDG values were employed 
 in earlier analyses. As a result, our numbers for the $\pi$ case are larger by $\sim 5\%$.
 \item We have not performed a dedicated study of the errors of the 
QCDf 
 contributions. Errors quoted in table \ref{Table1} are obtained rescaling the errors in 
 Refs.~\cite{ali} and \cite{bobeth2} according to the different central values obtained by them and us. As discussed extensively in these references (see also 
 \cite{Khod}), the dominating error for the $K$ case comes from the heavy-meson vector form factor, while in the $\pi$ case the choice of the renormalization 
 scale $\mu_b$ basically saturates the overall uncertainty (see, however, \cite{btokstar1}).
 \item Our study of the LD WA contributions to $B^-\to P^-\ell^+\ell^-$ has been performed with two different sets of form factors in the $(1,6)$ ($K$) 
 and $(1,8)$ ($\pi$) GeV$^2$ ranges. In the above-GeV intervals, the error has been estimated from the difference between these predictions. When including the 
 region immediately above threshold we have only employed the set of chiral-based form factors estimating the error as the difference between the results 
 obtained using dispersive form factors and a Guerrero-Pich-like resummation \cite{pionff}. Analogous procedure has been employed in order to obtain our 
 results for CP violation in the next section.
 \item The violations of lepton universality induced by kinematical effects on $R_K$ and $R_\pi$ are always given by the 
QCDf
 contribution. The $LD$ WA modification 
 is --in all the considered energy ranges-- smaller than the error of the 
QCDf 
 contribution. Therefore, $R_K=1.0003(1)$ in the $(1,6)$ GeV$^2$ range 
 \cite{bobeth} and $R_\pi=1.0006(1)$ in the $(1,8)$ GeV$^2$ range.
\end{itemize}
 
 When we add the contributions of $q^2$ values above 8 GeV$^2$, ref. \cite{ali}, our branching fraction corresponding to the full kinematical domain 
becomes $B(B^- \to \pi^- \ell^+\ell^-) \, = \, (2.6^{+0.4}_{-0.3})\times 10^{-8}$.

 These results can be compared to available data from Refs.~\cite{lhcb, lhcb2}:
\begin{eqnarray}
&&B(B^- \to \pi^- \mu^+\mu^-) = (2.3\pm 0.6\pm 0.1)\times 10^{-8}, \\
&&B(B^- \to K^- e^+e^-) = (1.56^{+0.20}_{-0.16} )\times 10^{-7}, \ \ \ {\rm for}\ {1<q^2<6\ {\rm GeV}^2}.
\end{eqnarray}
Our results and experimental data agree within error bars \footnote{Recent results reported in Ref. \cite{lhcb3} give a smaller BR for the muon case, but still consistent with the previous measurement.}. In the $B^- \to K^- \ell^+\ell^-$ decays, current errors on the (completely dominating) SD contribution 
do not allow to tell whether there is a tension between SM prediction and the LHCb measurement or not. With the smaller error expected on the branching fraction of 
the $B^- \to \pi^- \ell^+\ell^-$ decays from the next run of LHC measurements, one might be able to notice a tension between SM predictions and data. It must be 
noted that a reduction of the current error to less than a half will be able to pinpoint the LD WA contribution to these decays that we have been discussing.

\section{CP violation}
A direct CP asymmetry can be generated because the LD one-photon exchange and the 
QCDf 
 contributions to the amplitude of the $B^{\pm}\to P^\pm \ell^+\ell^-$ decays have different weak and strong phases, as it 
can be easily checked from eqs.~(\ref{eqQCDf}) and (\ref{eqLD}) \footnote{A possible large CP violation in the $B^{\pm}\to \pi^\pm \ell^+\ell^-$ decays induced 
by WA one-photon exchange contribution was proposed for the first time in ref.~\cite{Hou:2014dza} within the framework of QCDf.}. More precisely, the strong phase required to generate CP asymmetry in our case, arises from the 
LD WA contribution. It will dominate at low photon virtualities, since it stems from the finite decay widths of vector mesons that describe the electromagnetic structure of charged mesons. Thus, a (partially integrated over a finite $q^2$ range) CP asymmetry~\footnote{Other CP violating observables can be analyzed analogously.}
\begin{equation}
A_{CP}(P)=\frac{\Gamma(B^+\to P^+\ell^+\ell^-)-\Gamma(B^-\to P^-\ell^+\ell^-)}{\Gamma(B^+\to P^+\ell^+\ell^-)+\Gamma(B^-\to P^-\ell^+\ell^-)}\ ,
\end{equation}
can be generated from the interference of diagrams shown in Figure 1. By inserting the amplitudes of LD WA and 
QCDf
 contributions in the previous 
expressions, it can be shown that the width difference has the form (as seen in table \ref{Table1}, interferences are much smaller than the 
QCDf 
 and LD WA
contributions):
\begin{eqnarray}
\Delta_{CP} &=&\Gamma(B^+\to P^+\ell^+\ell^-)-\Gamma(B^-\to P^-\ell^+\ell^-) \nonumber \\ &=&
-32\alpha^2G_F^2f_Pf_B {\rm Im} \left\lbrace V_{tb}V^*_{tD}V^*_{ub}V_{uD}\right\rbrace
 \nonumber \\ 
&&\times \, \int dq^2\int ds_{12}
\frac{1}{q^2(M_B^2-m_P^2)}  \left[2(P_B\cdot P_+)(P_B\cdot P_-)-\frac{M_B^2q^2}{2}\right]  \\
&& \times {\rm Im} \left\lbrace \xi_P(q^2) F_V(q^2)\left[M_B^2 \left(F_P(q^2)-1\right)- 
m_P^2\left(F_B(q^2)-1\right)\right]\right\rbrace \, , \nonumber 
\end{eqnarray}
where $s_{12}=(p_K+p_+)^2$.

According to the matching of the LD and 
QCDf 
descriptions of the WA contributions, the latter will also violate CP for $q^2>2$ GeV$^2$, arising from on-shell radiating light-quarks. Finally, there will also be another source of 
QCDf 
 CP violation from light $q\bar{q}$ in loops \cite{Hou:2014dza}.
Taking all of them into account it  leads to:
\begin{eqnarray}
A_{CP} (P)= \left\{ \begin{array}{l} 
(16.1\pm 1.9)\%, \ \ \ \ {\rm for}\ P=\pi, \ 0.05 \leq q^2 \leq 8\ {\rm GeV}^2, \\ 
(7.8 \pm 2.9)\%, \ \ \ \ {\rm for}\ P=\pi, \ 1 \leq q^2 \leq 8\ {\rm GeV}^2, \\
(-1.0\pm 0.3)\% , \ \ \ \ {\rm for}\ P=K, \ 1 \leq q^2 \leq 6\ {\rm GeV}^2 .\\ \end{array} \right.
\end{eqnarray}
In the first energy range for the $\pi$ meson case, $83\%$ of $A_{CP}(\pi)$ has a LD WA origin, while this reduces to $31\%$ in the second.
In the case of kaons, for the $0.05 \leq q^2 \leq 6$ GeV$^2$ range, the amount of CP violation basically doubles but is completely dominated by LD WA contributions (which is already $70\%$ in the $1 \leq q^2 \leq 6$ GeV$^2$ interval). The quoted error in our results stems from the systematic error attributed to parametrizations of the light meson electromagnetic form factors.

In ref.~\cite{Hou:2014dza}, using QCDf, the $A_{CP} (\pi)$ asymmetry was predicted to be larger than ours (for comparison, see Table \ref{Table2} for three ranges of the lepton-pair invariant mass), while the results in 
the recent paper \cite{Hambrock:2015wka} lie somehow in between of both predictions. Note that in the QCDf approach the strong phase for low values of $q^2$ is dominated by the photon emission off the spectator $u$ quark (in the $B$ meson) in the WA diagram, which may be on-shell after having emitted the photon (for larger values of $q^2$, a smaller strong CP phase arises from light $q\bar{q}$ in loops from form factor $B\to \pi$ contributions). We note that this CP violating effect is maximal towards the $q^2$ threshold, where the applicability of QCDf is more questionable. Measurements of binned CP asymmetries in this and the $B^\pm \to V^\pm \ell^+ \ell^-$ ($V=\rho,K^*$) decays may be sensitive to the modelization of the WA contributions at long distances.

We note that the signs of the results in refs.~\cite{Hou:2014dza, Hambrock:2015wka} have been switched in Table \ref{Table2} because our convention for defining $A_{CP} (\pi)$ is opposite. Within our approach, the solely contribution of the 
B-meson electromagnetic form factor to the CP asymmetry is completely negligible, at the $\mathcal{O}(10^{-4})$ level.

\begin{table}[!t]
\caption{\small Our results for $A_{CP} (\pi)$ (in $\%$) are compared to those in ref.~\cite{Hou:2014dza} for different energy ranges. We note that in the $(1,6)$ GeV$^2$ range the recent result in Ref.~\cite{Hambrock:2015wka}, 
$(14.3^{+3.5}_{-2.9})\%$ is in agreement with both determinations, although closer to ref.~\cite{Hou:2014dza}.}
\begin{tabular}{|c|c|c|}
\hline
$(q^2_{min},q^2_{max})$ &Ref.~\cite{Hou:2014dza}&Our results\\
\hline
\ \ $(1,8)$ GeV$^2$ \ \ & \ \ \ $13\pm 2$\ \ \ & \ \ \ $7.8\pm2.9$ \ \ \\\
\ \ $(1,6)$ GeV$^2$ \ \ & \ \ \ $16\pm2$ \ \ \ & \ \ \ $9.2\pm1.7$ \ \ \ \\
\ \ $(2,6)$ GeV$^2$ \ \ & \ \ \ $13^{+2}_{-3}$ \ \ \ & \ \ \ $7.7\pm0.5$\ \ \ \\
\hline
\end{tabular}\label{Table2}
\end{table}

 This $\mathcal{O}({\rm few}\ \%)$ CP violating figures shall enhance the case for their measurements. Current values at the PDG 
 are well compatible with zero \cite{pdg2014} in $B^\pm\to K^\pm\ell^+\ell^-$ (and also in $B^\pm\to K^{*\pm}\ell^+\ell^-$) decays, while this observable is not 
 yet reported in the $B^\pm\to \pi^\pm\ell^+\ell^-$ case. The most recent measurement reported by the LHCb collaboration for the integrated CP asymmetry,  namely $A_{CP}=0.11\pm 0.12\pm 0.01$ \cite{lhcb3}, is still consistent with the different theoretical predictions. 

Therefore, despite the fact that the tree-level one-photon exchange diagrams give a small contribution to the decay rates (specially for the $K$ case), they can 
generate a non-negligible CP asymmetry within the Standard Model. This CP asymmetry altogether with measurements of the decay rates can be used as a test of 
New Physics in the rare $B^{\pm} \to P^{\pm}\ell^+\ell$ decays. This makes us emphasize the need of a dedicated measurement of these observables in the next LHC 
run at LHCb and in the forthcoming Belle-II experiment.

\section{Conclusions} \label{conclusions}

In this paper we have considered the one-photon exchange contribution to the rare $B^{\pm} \to P^{\pm}\ell^+\ell^-$ decays, with $P=\pi$ or $K$. Its effects in 
the decay rates of the $P=K$ case turn out to be of order 1$\%$ with respect to the (top quark loop dominated) SD contribution for the range 
$1\leq q^2\leq 6$ GeV$^2$ of the squared lepton pair invariant mass. We do not foresee forthcoming measurements being sensitive to this contribution in the near 
future. On the contrary, this fact confirms the suitability of this range for new physics searches.

In the case of a $\pi^{\pm}$ meson in the final state, the corresponding effect turns out to be significant in integrated observables starting close to threshold. 
This suggests to take --in analogy to the case of a final state with $K$ -- the range $1\leq q^2\leq 8$ GeV$^2$ for precision measurements, since the LD WA contribution is 
reduced to less than 10$\%$ with a negligible uncertainty in that interval. On the other hand, more refined measurements of the fully integrated branching fraction 
for this decay could be sensitive to our contribution once the error is reduced below a half of the current uncertainty.

Interestingly, the different weak and strong phases of the  
QCDf
 and LD WA (one-photon exchange) contributions are capable to generate a CP asymmetry. Again, this CP asymmetry is large in the case of a pion in the final state for $0.05 \leq q^2\leq 8$ GeV$^2$ values of the lepton-pair invariant mass, but also 
sizable and worth to measure in the $1 \leq q^2\leq 8$ GeV$^2$ interval. For the kaon case, the range $1 \leq q^2\leq 6$ GeV$^2$ is optimal for such a search. Our CP violation results are smaller than those obtained within QCDf because of the different description of the WA amplitudes at low energies. Future measurements shall be sensitive to this kind of contribution and shed light on its appropriate description. More refined measurements of this CP asymmetry and of the magnitudes of the decay rates at LHCb and future $B$-superfactories can provide another non-trivial test of the Standard Model or may furnish indications of New Physics.

\medskip

{\bf Acknowledgements}: The authors would like to thank Conacyt (M\'exico) 
for financial support. We appreciate very much helpful correspondence with Bogdan Malaescu concerning the BaBar parametrization of 
the kaon electromagnetic form factor. We thank Javier Virto for useful discussions on the topic of this article.


\end{document}